\documentclass{natureprintstylev4}
\usepackage{graphicx}
\usepackage{epsfig}
\usepackage{amsmath,amssymb,amsfonts}
\usepackage{mathtools}
\usepackage{booktabs}
\usepackage{multirow}
\usepackage{longtable}
\usepackage{textcomp}
\usepackage[utf8]{inputenc}
\usepackage[T1]{fontenc}
\usepackage[svgnames]{xcolor}
\usepackage{hyperref}
\usepackage[numbers,sort&compress]{natbib}
\usepackage{astjnlabbrev-nature}

\DeclareUnicodeCharacter{223C}{$\sim$}   % ∼
\DeclareUnicodeCharacter{03BC}{$\mu$}    % μ
\DeclareUnicodeCharacter{2273}{$\gtrsim$} % ≳
\DeclareUnicodeCharacter{03B1}{$\alpha$} % α
\DeclareUnicodeCharacter{03B2}{$\beta$}  % β
\DeclareUnicodeCharacter{2272}{$\lesssim$}
\DeclareUnicodeCharacter{2248}{$\approx$}
\DeclareUnicodeCharacter{2264}{$\leq$}

\usepackage{hyperref}
\hypersetup{
   colorlinks=true,
   linkcolor=blue,
   filecolor=green,
   citecolor=blue,
   urlcolor=blue
}

\newcommand{\rev}[1]{{#1}}

\let\cline\cmidrule

\begin{document}

\title{Early metal-enriched baryon cycling before the midpoint of cosmic reionization}

\author{
Yongda Zhu$^{1,*}$,
Zhiyuan Ji$^{1,*}$,
George D. Becker$^{2}$,
Jiani Ding$^{1}$,
Eiichi Egami$^{1}$,
Xiaohui Fan$^{1}$,
Xiangyu Jin$^{3}$,
Weizhe Liu$^{1}$,
Jianwei Lyu$^{1}$,
Zheng Ma$^{1}$,
Suprabhas Narisetty$^{1}$,
George H. Rieke$^{1}$,
Yunjing Wu$^{4,1}$,
Minghao Yue$^{1}$,
Junyu Zhang$^{1}$,
Marcia J. Rieke$^{1}$
}

\maketitle

\begin{affiliations}
\item Steward Observatory, University of Arizona, 933 North Cherry Avenue, Tucson, AZ 85721, USA
\item Department of Physics \& Astronomy, University of California, 900 University Avenue, Riverside, CA 92521, USA
\item Department of Astronomy, University of Michigan, 1085 S. University Ave., Ann Arbor, MI 48109, USA
\item Kavli Institute for the Physics and Mathematics of the Universe (WPI), The University of Tokyo Institutes for Advanced Study, The University of Tokyo, Kashiwa, Chiba 277-8583, Japan
\end{affiliations}
$^\ast$Correspondence and requests for materials should be addressed to Yongda Zhu and Zhiyuan Ji: \mbox{yongdaz@arizona.edu}, \mbox{zhiyuanji@arizona.edu}.

~

\begin{abstract}
Models predict that chemical enrichment and gas redistribution should begin rapidly once star formation starts, but direct constraints at the earliest epochs have been scarce. Here we show that metal-enriched gas in multiple ionic phases was already present around galaxies before the midpoint of cosmic reionization. Using JWST/NIRSpec rest-frame ultraviolet spectroscopy from SPURS, we detect blueshifted metal absorption in three galaxies at $7.2<z<9.3$. The detected transitions span neutral, low-ionization, and high-ionization species, including O\,\textsc{i}, Si\,\textsc{ii}, C\,\textsc{ii}, Si\,\textsc{iv}, and C\,\textsc{iv}, with velocity offsets of $|\Delta v|\sim 50$--$250\,\mathrm{km\,s^{-1}}$ relative to nebular systemic redshifts. The ionic coexistence, overlapping velocity structure, and equivalent-width ratios are consistent with outflowing or otherwise kinematically disturbed galaxy-associated gas, implying rapid metal enrichment. These results show that key conditions for baryon cycling were established in at least a subset of luminous galaxies within the first several hundred million years of cosmic time, well before the completion of reionization.
\end{abstract}

\section*{Introduction}

Galaxies influence their own growth and their surrounding environments by internally circulating gas, metals, and energy, and by exchanging them with their circumgalactic medium and neighboring systems; these processes are collectively referred to as the baryon cycle \cite[e.g.,][]{tumlinson_circumgalactic_2017}. 
Recent observations have pushed the emergence of luminous galaxies to increasingly early cosmic times, including spectroscopically confirmed galaxies at $z>14$ \cite{carniani_shining_2024,naidu_cosmic_2025} and Ly$\alpha$ emitters at $z>8$ \citep[e.g.,][]{chen_impact_2025,witstok_witnessing_2025}, demonstrating that significant star formation was already underway during the earliest phases of cosmic reionization. Despite this progress, it remains unclear when chemically enriched gas first appeared around galaxies. In particular, it is unknown whether metal-enriched gas in multiple ionic phases was already present around galaxies before the midpoint of reionization at $z\sim7$--$8$ (cosmic time $\sim0.65$--$0.75$~Gyr) \citep[e.g.,][]{mason_inferences_2019,planck_collaboration_planck_2020}, well prior to the end of reionization at $z\sim5.3$ \citep[e.g.,][]{zhu_probing_2023}. Theoretical models predict that metal production and stellar feedback can rapidly enrich gas associated with early galaxies and generate complex ionization structures
\citep[e.g.,][]{muratov_gusty_2015,kim_high-redshift_2019,pandya_characterizing_2021}, but direct observational evidence for such enriched gas at these earliest epochs has remained scarce.

During the epoch of reionization, metal enrichment beyond galaxies has been measured primarily at $z\lesssim6.5$ through absorption systems detected along quasar sightlines \citep[e.g.,][]{fan_quasars_2023}. These measurements probe largely random intergalactic paths and are only indirectly connected to individual galaxies.
Direct probes of gas in and around galaxies during reionization have largely relied on rest-frame optical emission lines and broadband spectral energy distributions (SEDs), which primarily trace internal galaxy properties \citep[e.g.,][]{stark_galaxies_2016}. Recent observations have pushed such measurements to very early times, including detections of strong metal emission lines (e.g., [O\,III]) in galaxies up to $z\sim14$, demonstrating that substantial metal enrichment can already occur within the interstellar media of the first galaxies \citep[e.g.,][]{finkelstein_complete_2023}. 
Alternative background sources for absorption studies, such as gamma-ray bursts or strongly lensed systems, remain rare and are limited to small samples or lower redshifts \citep[e.g.,][]{gatkine_cgm-grb_2022}. Recent stacking analyses of galaxy spectra have provided indirect evidence for average absorption signatures at high redshift \citep[e.g.,][]{glazer_stacking_2025}, but direct, system-by-system constraints on enriched gas associated with galaxies during reionization have remained scarce.

Medium-resolution ($R\sim1000$) near-infrared spectroscopy with NIRSpec on the \textit{James Webb Space Telescope} (JWST), together with deep integrations reaching high continuum signal-to-noise ratios in the rest-frame ultraviolet, now enables absorption-line measurements directly against the stellar continua of individual galaxies at $z>7$, a regime only recently accessible to such observations. This capability allows metal absorption to be measured in a velocity frame physically tied to the galaxies themselves and enables empirical constraints on the ionization structure using species that trace neutral, low-ionization, and high-ionization gas. In this work, we use these observations to place direct constraints on the presence and ionization structure of chemically enriched, galaxy-associated gas at $z\simeq7$--9.

\begin{figure*}[!ht]
\centering
\includegraphics[width=\linewidth]{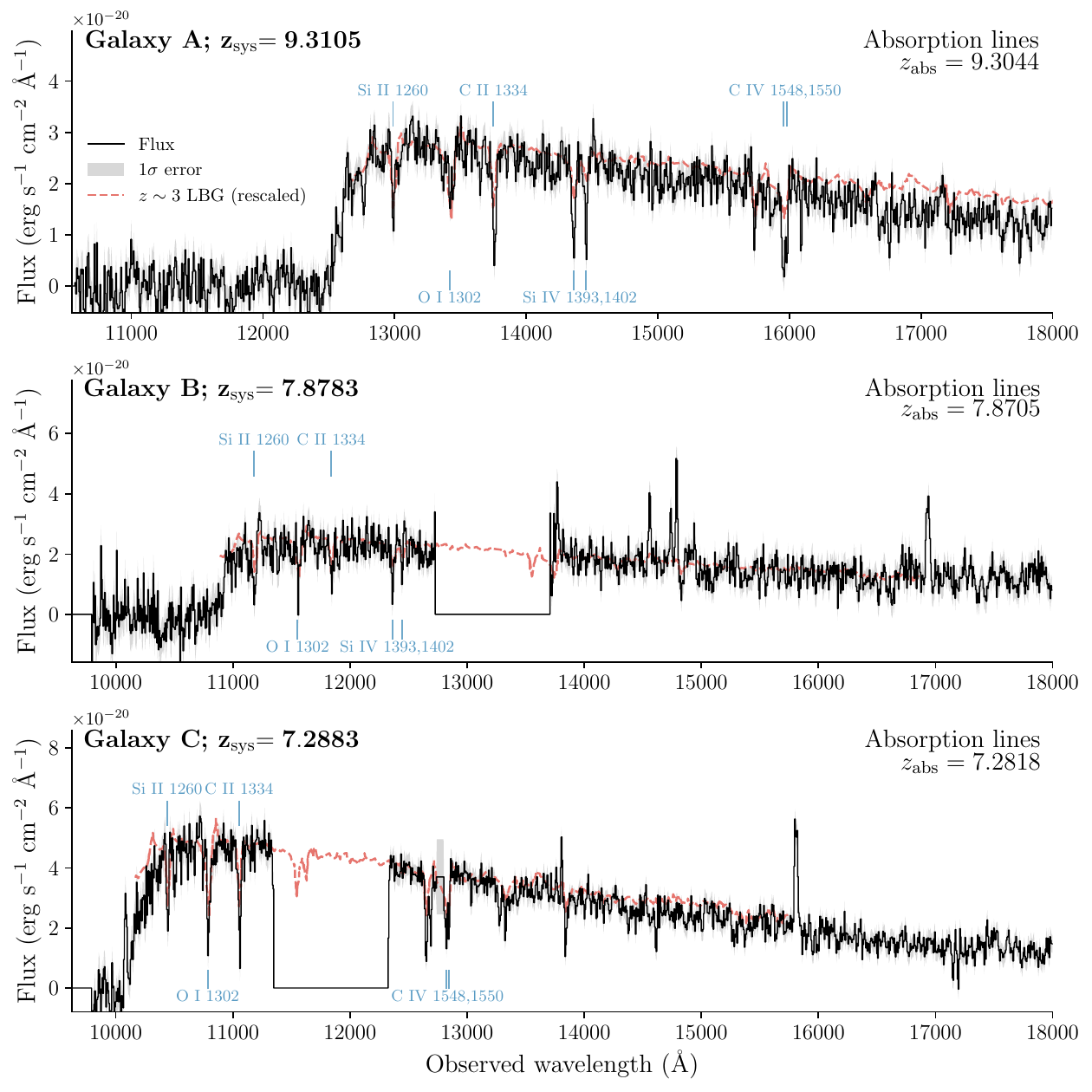}
\caption{Spectra covering the rest-frame ultraviolet of the three galaxies used in this work, shown in observed wavelength (Galaxy~A at $z_{\rm sys}=9.3105$, Galaxy~B at $z_{\rm sys}=7.8783$, and Galaxy~C at $z_{\rm sys}=7.2883$). Black curves show the observed JWST/NIRSpec spectra with gray shading indicating $1\sigma$ uncertainties. Vertical ticks mark the expected positions of neutral, low-ionization, and high-ionization metal absorption lines at the absorber redshift $z_{\rm abs}$. For reference, the red dashed curve shows a composite $z\!\sim\!3$ Lyman-break galaxy (LBG) spectrum \cite{shapley_rest-frame_2003}, normalized to match the median continuum flux near rest-frame 1450\,\AA. The spectra show metal absorption features at $z\sim 7-9$ broadly similar to lower-redshift galaxy spectra.}
\label{fig:spec_uv}
\end{figure*}

\begin{figure*}[!ht]
\centering
\includegraphics[width=\textwidth]{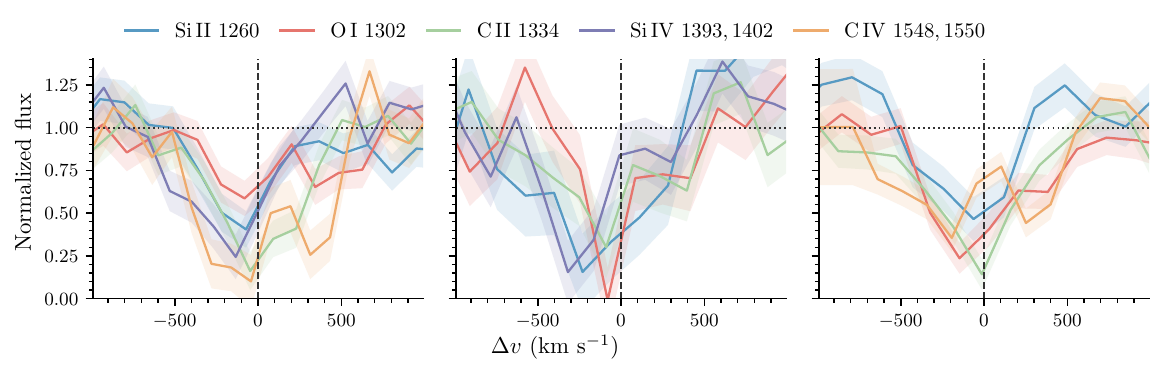}
\caption{Continuum-normalized absorption-line profiles plotted as a function of velocity relative to the galaxy systemic redshift for Galaxy~A (left), Galaxy~B (middle), and Galaxy~C (right). Shaded regions indicate $1\sigma$ uncertainties.}
\label{fig:vel_stack}
\end{figure*}

\begin{figure*}[!ht]
\centering
\includegraphics[width=1\textwidth]{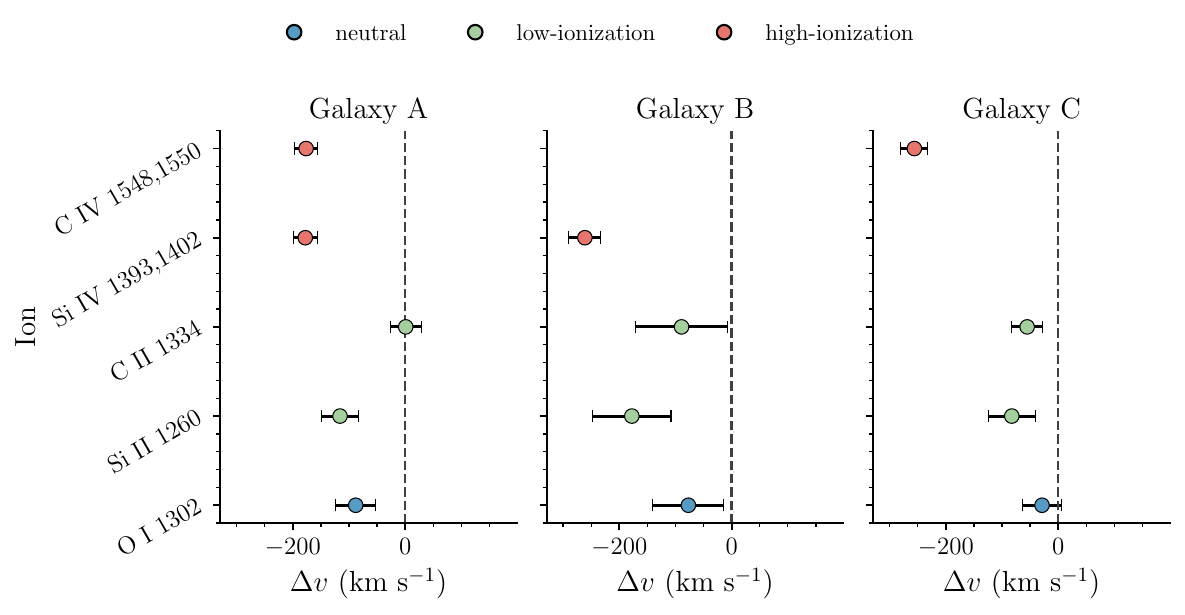}
\caption{
\rev{Ionization-dependent absorption-line velocity structure. Points show the fitted velocity offsets of individual absorption features relative to the systemic redshift defined by [O\,III] $\lambda5008$. Colors denote neutral, low-ionization, and high-ionization species. Error bars show the $1\sigma$ centroid uncertainties from the absorption-line fits.}
}
\label{fig:ion_velocity}
\end{figure*}

\begin{figure*}[!ht]
\centering
\includegraphics[width=0.65\linewidth]{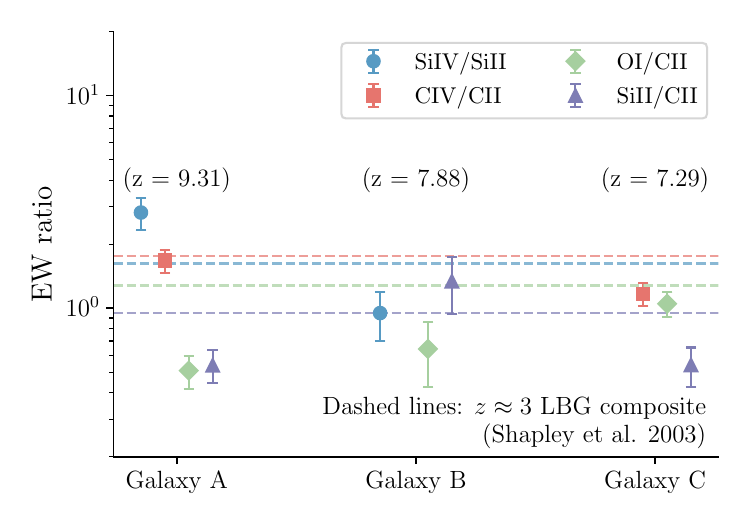}
\caption{
Major empirical rest-frame equivalent-width ratios for metal absorption lines in the three galaxies. Shown are ratios between high- and low-ionization species (Si\,IV/Si\,II and C\,IV/C\,II) and ratios involving neutral and low-ionization species (Si\,II/C\,II and O\,I/C\,II), computed using the model equivalent widths measured from the NIRSpec spectra. These ratios provide empirical diagnostics of the relative ionization structure of metal-bearing gas. Dashed horizontal lines indicate reference values constructed from the $z\!\sim\!3$ LBG composite spectrum in \cite{shapley_rest-frame_2003}. We note that individual lower-redshift galaxies exhibit substantial intrinsic scatter in these ratios ($\sim$ a factor of two; \citep[e.g.,][]{chisholm_shining_2016}); the composite therefore provides a representative reference level.
}
\label{fig:EWR}
\end{figure*}

\section*{Results} 
Continuum absorption-line measurements at $z>7$ require ultraviolet-bright galaxies with sufficiently high signal-to-noise rest-frame ultraviolet spectra. In the currently available JWST data, this criterion is met by three galaxies in the Cycle~4 program SPectroscopic Ultra-deep Reionization-era Survey (SPURS; PID~9214, PIs: Mason, Stark; \citep{chen_spurs_2026}), all of which have rest-frame ultraviolet continuum detected near 1450\,\AA\ at $S/N>10$ and $M_{\rm UV}<-20.9$. These systems were selected on the basis of continuum detectability, without foreknowledge of any metal absorption feature or metallicity. The SPURS observations provide deep, medium-resolution NIRSpec spectroscopy, enabling absorption-line measurements directly against the stellar continuum. Figure~\ref{fig:spec_uv} presents the observed spectra covering the rest-frame ultraviolet for all three galaxies.
SED modeling of the galaxies' HST and JWST photometry provides additional host-galaxy properties. The three systems have stellar masses of $\log(M_\star/M_\odot)=9.80$, 9.11, and 9.48, and blue observed ultraviolet slopes of $\beta_{\rm UV,obs}=-2.19$, $-2.12$, and $-2.26$ for Galaxies~A, B, and C, respectively. A detailed analysis of the stellar populations, nebular emission, and feedback properties of these galaxies is presented in complementary works (\citep{chen_spurs_2026}; \citep{keerthi_vasan_g_jwst_2026}).

For the highest-redshift system (Galaxy~A, $z_{\rm sys}=9.3105$), we adopt the systemic redshift from [O\,III] $\lambda5008$. Although this line lies near the red end of the G395M coverage, the fitted line window is not truncated by the data boundary, and the derived centroid is consistent with the other detected nebular features within the measurement uncertainties. We identify absorption from neutral gas (O\,I $\lambda1302$), low-ionization gas (Si\,II $\lambda1260$ and C\,II $\lambda1334$), and high-ionization gas (Si\,IV $\lambda\lambda1393,1402$ and C\,IV $\lambda\lambda1548,1550$). 
We next assess whether such galaxy-associated metal absorption is unique to a single object. We apply the same analysis to the two other continuum-detected galaxies in the program, Galaxies~B and~C, at $z_{\rm sys}=7.8783$ and $z_{\rm sys}=7.2883$, respectively. In both systems, we detect the same suite of neutral and low-ionization metal absorption tracers as in Galaxy~A, including O\,I, Si\,II, and C\,II. 

The absorption features in all three galaxies are broadly similar in appearance to those seen in lower-redshift star-forming galaxies, as illustrated by comparison with a composite $z\!\sim\!3$ Lyman-break galaxy (LBG) spectrum in \cite{shapley_rest-frame_2003}, which we use as a well-studied empirical reference for rest-frame ultraviolet absorption in star-forming galaxies.
We note that the composite spectrum, constructed from many galaxies with differing velocity offsets and intrinsic line widths, naturally produces absorption features that are broader and shallower than those in individual systems. In the remainder of this section, we focus on empirical constraints from the absorption-line kinematics and relative equivalent widths, using these measurements to probe the ionization state and physical conditions of metal-enriched, galaxy-associated gas during the early phases of baryon cycling.

\begingroup
% \small
% \setlength{\tabcolsep}{20pt}
\begin{table*}[!ht]
    \centering
    % \hspace{-1cm}
    % \fontsize{12pt}{12pt}\selectfont
    \small
\begin{tabular}{llccc}
\toprule
 &  & Galaxy A & Galaxy B & Galaxy C \\
\midrule
Systemic redshift &  & 9.3105 $\pm$ 0.0003 & 7.8783 $\pm$ 0.0002 & 7.2883 $\pm$ 0.0002 \\
\cline{1-5}
$v$ (km s$^{-1}$) & O\,I 1302 & -88 $\pm$ 35 & -77 $\pm$ 63 & -29 $\pm$ 34 \\
$W_0$ (\AA) & O\,I 1302 & 1.00 $\pm$ 0.16 & 1.11 $\pm$ 0.24 & 2.11 $\pm$ 0.22 \\
$W_{0,\rm model}$ (\AA) & O\,I 1302 & 1.03 $\pm$ 0.15 & 1.11 $\pm$ 0.28 & 2.14 $\pm$ 0.21 \\
$\tau_0$ & O\,I 1302 & 8.10 & 50.00 & 25.80 \\
$\min(F/F_c)$ & O\,I 1302 & 0.601 & -0.010 & 0.236 \\
\cline{1-5}
$v$ (km s$^{-1}$) & Si\,II 1260 & -116 $\pm$ 33 & -178 $\pm$ 70 & -83 $\pm$ 42 \\
$W_0$ (\AA) & Si\,II 1260 & 1.08 $\pm$ 0.17 & 2.20 $\pm$ 0.47 & 1.08 $\pm$ 0.22 \\
$W_{0,\rm model}$ (\AA) & Si\,II 1260 & 1.09 $\pm$ 0.16 & 2.30 $\pm$ 0.45 & 1.10 $\pm$ 0.21 \\
$\tau_0$ & Si\,II 1260 & 15.05 & 31.60 & 12.08 \\
$\min(F/F_c)$ & Si\,II 1260 & 0.404 & 0.156 & 0.465 \\
\cline{1-5}
$v$ (km s$^{-1}$) & C\,II 1334 & 1 $\pm$ 28 & -89 $\pm$ 82 & -55 $\pm$ 28 \\
$W_0$ (\AA) & C\,II 1334 & 1.98 $\pm$ 0.20 & 1.65 $\pm$ 0.40 & 2.10 $\pm$ 0.19 \\
$W_{0,\rm model}$ (\AA) & C\,II 1334 & 2.02 $\pm$ 0.19 & 1.72 $\pm$ 0.39 & 2.04 $\pm$ 0.18 \\
$\tau_0$ & C\,II 1334 & 32.89 & 19.65 & 25.54 \\
$\min(F/F_c)$ & C\,II 1334 & 0.161 & 0.301 & 0.142 \\
\cline{1-5}
$v$ (km s$^{-1}$) & Si\,IV 1393,1402 & -178 $\pm$ 21 & -261 $\pm$ 28 & \dots \\
$W_0$ (\AA) & Si\,IV 1393,1402 & 3.06 $\pm$ 0.30 & 1.94 $\pm$ 0.39 & \dots \\
$W_{0,\rm model}$ (\AA) & Si\,IV 1393,1402 & 3.07 $\pm$ 0.28 & 2.18 $\pm$ 0.36 & \dots \\
$\tau_0$ & Si\,IV 1393,1402 & 31.65 & 19.65 & \dots \\
$\min(F/F_c)$ & Si\,IV 1393,1402 & 0.232 & 0.155 & \dots \\
\cline{1-5}
$v$ (km s$^{-1}$) & C\,IV 1548,1550 & -176 $\pm$ 21 & \dots & -236 $\pm$ 21 \\
$W_0$ (\AA) & C\,IV 1548,1550 & 3.11 $\pm$ 0.30 & \dots & 2.06 $\pm$ 0.20 \\
$W_{0,\rm model}$ (\AA) & C\,IV 1548,1550 & 3.38 $\pm$ 0.28 & \dots & 2.03 $\pm$ 0.19 \\
$\tau_0$ & C\,IV 1548,1550 & 43.53 & \dots & 15.01 \\
$\min(F/F_c)$ & C\,IV 1548,1550 & 0.101 & \dots & 0.383 \\
\cline{1-5}
\bottomrule
\end{tabular}
\caption{
Absorption-line measurements for galaxy-associated metals at $z\simeq7$--9. Systemic redshifts are derived from rest-frame optical emission lines, using [O\,\textsc{iii}] $\lambda5008$ as the adopted reference, and are reported with their $1\sigma$ statistical uncertainties. Velocity offsets $v$ are measured relative to the adopted systemic redshift and are quoted with their $1\sigma$ statistical uncertainties. Rest-frame equivalent widths $W_0$ are reported for single transitions and as summed values for doublets. The model equivalent widths are derived from Voigt-profile fits. Optical-depth normalizations $\tau_0$ are reported for completeness only, as saturation and partial covering may be present and some values reach fitting bounds; our conclusions do not rely on $\tau_0$ or Voigt-derived column densities. The minimum normalized flux $\min(F/F_c)$ provides a model-independent measure of line depth. Negative $\min(F/F_c)$ values arise from noise fluctuations and do not imply physical flux below zero. Uncertainties correspond to $1\sigma$ statistical errors. 
}
\label{tab:ed_abs}
\end{table*}
\endgroup

To assess whether the detected ions trace a common absorber component, we compare their kinematics in the systemic velocity frame. Figure~\ref{fig:vel_stack} shows the continuum-normalized absorption profiles of different ionic species for each galaxy, plotted as a function of velocity relative to the systemic redshift. Neutral (O\,I), low-ionization (Si\,II, C\,II), and high-ionization (Si\,IV, C\,IV) transitions show broadly similar, overlapping velocity structure, with a common bulk blueshift from systemic and comparable velocity extents, typically spanning $\sim -200\,\mathrm{km\,s^{-1}}$. We use ``multiphase'' here in an observational sense, referring to the simultaneous detection of neutral, low-ionization, and high-ionization absorption within overlapping velocity intervals. The individual centroids listed in Table~\ref{tab:ed_abs} differ, and the dominant pattern is a common bulk blueshift. The overlapping absorption structure across ions spanning more than two orders of magnitude in ionization potential suggests that the absorption arises from galaxy-associated gas with overlapping cross-ion kinematics. Velocity information alone does not uniquely distinguish between compact outflowing material and gas distributed over larger circumgalactic scales, as both scenarios can produce blueshifted absorption depending on geometry and driving mechanism. However, the observed overlapping velocity structure across multiple ions indicates that the gas is dynamically associated with the host galaxy; these bulk blueshifts cannot arise from Hubble expansion relative to the galaxy rest frame. Multiple metal absorbers are detected at comparable velocity offsets from systemic in all continuum-detected targets, indicating that kinematically disturbed, metal-enriched gas was already present in multiple galaxies during the reionization era. \rev{Figure~\ref{fig:ion_velocity} summarizes the fitted velocity centroids by ionization class.}

The velocity widths of the detected metal absorption lines provide an additional empirical constraint on the physical state of the absorbing gas. Assuming purely thermal broadening, we derive upper limits on the gas temperature from the fitted line widths (see \textit{Methods}). These limits are weak and exceed the ionization survival temperatures of the detected species, indicating that non-thermal motions must dominate the observed profiles. At the present spectral resolution, the fitted line widths are best interpreted as effective measures of unresolved absorption structure instead of as precise intrinsic dispersions for individual ions. The evidence for common kinematics comes from the shared bulk blueshift and overlapping velocity extent across ions.

Finally, we characterize the ionization structure of the absorbing gas using empirical rest-frame equivalent-width ratios. Figure~\ref{fig:EWR} compares ratios between high- and low-ionization species (Si\,IV/Si\,II and C\,IV/C\,II) and ratios involving neutral and low-ionization species (Si\,II/C\,II and O\,I/C\,II) for all three galaxies. Galaxies~B and~C have broadly similar EW ratios and are also comparable to the $z\!\sim\!3$ LBG composite reference. Galaxy~A spans a wider region of EW-ratio space, with lower O\,I/C\,II and Si\,II/C\,II ratios and a higher Si\,IV/Si\,II ratio. Lower-redshift samples show that Si-based equivalent-width ratios have relatively low dispersion, whereas ratios involving O\,I vary by about a factor of two at fixed stellar mass \citep[e.g.,][]{chisholm_shining_2016}. In this sense, the range spanned by these three galaxies does not obviously exceed the level of scatter reported at lower redshift, while still showing clear system-to-system variation at high redshift.

\section*{Discussion}

\subsection*{What the observations require, and what they do not.}
Our central results are that metal absorption is measured in a systemic frame defined by rest-frame optical nebular lines, that the absorption is blueshifted by $\sim50$--250\,km\,s$^{-1}$, and that neutral, low-ionization, and high-ionization absorption show broadly overlapping velocity structure across the sample. The available nebular lines support the adopted systemic redshifts: using H$\beta$ instead of [O\,III] $\lambda5008$ where detected does not remove the blueshifted absorption signature.

Such offsets and overlapping cross-ion velocity structure are commonly observed in lower-redshift galaxies and suggest that the absorption arises in galaxy-associated gas with multiple ionic phases, disfavoring a single quiescent component \citep[e.g.,][]{werk_cos-halos_2014,davies_jwst_2024}. The fitted centroids may also indicate that the high-ionization absorption is more blueshifted than the neutral and low-ionization absorption in the current sample: by $\sim60$--180\,km\,s$^{-1}$ in Galaxy~A, $\sim80$--180\,km\,s$^{-1}$ in Galaxy~B, and $\sim150$--210\,km\,s$^{-1}$ in Galaxy~C, depending on the comparison line \rev{(Fig.~\ref{fig:ion_velocity})}. Higher-resolution spectra would be needed to establish a detailed ionization-dependent velocity structure. \rev{This velocity ordering is consistent with simulations in which more highly ionized material traces a faster or more extended component of a multiphase outflow \citep[e.g.,][]{li_prevalence_2026}, and provides an empirical signature of ionization-dependent kinematic stratification. It may reflect acceleration of ionized gas, cool clouds embedded in a faster medium, or an increasing ionization state along the flow.} 

Our interpretation rests on the empirical combination of overlapping absorption across ions spanning a wide range of ionization potentials and the simultaneous presence of blueshifted low- and high-ionization absorption, which is naturally produced by gas with multiple ionic phases in star-forming systems, including outflowing interstellar material and inner circumgalactic structures \citep[e.g.,][]{oppenheimer_multiphase_2018,veilleux_cool_2020}. \rev{This interpretation is consistent with zoom-in simulations such as FIRE-2, in which stellar feedback drives multiphase galactic winds with phase-dependent mass, momentum, energy, and metal loading across cold, warm, and hot gas \citep{pandya_characterizing_2021}.} \rev{The observed $\sim50$--250\,km\,s$^{-1}$ blueshifts at $z\simeq7$--9 are therefore consistent with early feedback-driven redistribution of metal-bearing gas into multiple ionization phases.} Even allowing for an interstellar contribution to some low-ion absorption, the simultaneous presence of overlapping high-ionization absorption requires gas phases not confined to a cold ISM. Comparisons in ion-ratio space are consistent with well-studied absorbers at $z\sim2$--3 \citep[e.g.,][]{prochaska_cos-halos_2017} and provide an empirical reference.

Still, regardless of whether individual components arise in outflowing or recycled material, the observations require chemically enriched gas with disturbed kinematics. \rev{They do not require a unique acceleration model, but they do require early redistribution of metals by feedback or related baryon-cycle processes.} At the same time, the broader high-redshift absorber population exhibits substantial diversity in metallicity, ranging from chemically enriched systems to extremely metal-poor absorbers \citep[e.g.,][]{simcoe_extremely_2012,davies_xqr-30_2023}, indicating that early chemical enrichment proceeded in a spatially and temporally inhomogeneous manner.

% \subsection*{Kinematically overlapping, multiphase absorbers.}
\subsection*{Overlapping cross-ion kinematics in multiphase absorbers.}
The broadly overlapping absorption profiles indicate a common cross-ion velocity structure. In particular, the presence of C\,II and C\,IV within the same blueshifted absorption complexes is difficult to reconcile with a single quiescent interstellar component, and instead points to galaxy-associated gas containing both low- and high-ionization material. The detection of C\,IV at $z\gtrsim7$ is itself notable, given the reported decline of strong C\,IV absorption in quasar sightlines with increasing redshift \citep[e.g.,][]{fan_quasars_2023}. This contrast may reflect the fact that C\,IV is more readily detected in gas closely associated with galaxies \citep[e.g.,][]{lan_multi-phase_2025}. Overlapping absorption from multiple ionic phases is commonly observed in the circumgalactic and interstellar media of lower-redshift galaxies \citep[e.g.,][]{bordoloi_cos-dwarfs_2014,qu_cosmic_2023}. 
% and comparable blueshifted absorption has also been reported in gravitationally lensed star-forming systems at lower redshift \citep[e.g.,][]{quider_ultraviolet_2009,quider_study_2010}.
 A weak or obscured active galactic nucleus (AGN) contribution cannot be excluded in every case, but the present data do not show direct AGN signatures (e.g., broad emission lines), and the available nebular measurements do not require an AGN interpretation (see \textit{Methods}). The absorption is therefore most naturally described in terms of galaxy-associated gas traced simultaneously by neutral, low-ionization, and high-ionization species.

The observed line widths provide an independent and complementary constraint. Interpreting the fitted Gaussian core widths as arising purely from thermal motions yields strict upper limits of $T_{\max}\sim10^{7}$\,K (see \textit{Methods}). Because any contribution from turbulence, bulk flows, or unresolved velocity substructure would reduce the thermal component of the observed width, these values should be regarded as upper limits. Under this assumption, thermal broadening in a single static cold or warm ISM component cannot readily account for the observed line widths \citep{li_prevalence_2026}. These conclusions rely directly on the measured velocities, line widths, and ionic coexistence, and do not require detailed ionization modeling or absolute metallicity estimates.

While the physical extent of the absorbing gas cannot be spatially resolved, the coexistence of neutral, low-ionization, and high-ionization species within a shared blueshifted absorption complex indicates a multiphase absorbing structure associated with the galaxy. Such configurations are common in star-forming systems and may arise in interstellar outflows, inner circumgalactic gas, or a combination of both. The present data do not uniquely distinguish between these possibilities.

\subsection*{Ionization structure and EW-ratio diversity.}
We use rest-frame equivalent-width ratios to characterize the relative ionization structure of the absorbing gas. This approach follows established studies of star-forming galaxies at lower redshift, where EW ratios between low- and high-ionization species have been widely used as observational diagnostics of multiphase gas and outflows without relying on detailed ionization modeling \citep[e.g.,][]{shapley_rest-frame_2003,du_redshift_2018}. Figure~\ref{fig:EWR} compares ratios between high- and low-ionization species (Si\,IV/Si\,II and C\,IV/C\,II) and ratios involving neutral and low-ionization species (Si\,II/C\,II and O\,I/C\,II) for all three galaxies. These ratios provide a largely model-independent view of the relative strengths of neutral, low-ionization, and high-ionization absorption, without requiring ionization corrections or assumptions about absolute metallicity.

Galaxies~B and~C ($z\simeq7.2$--7.8) occupy a relatively narrow region of EW-ratio space and are broadly consistent with each other, as well as with the reference values constructed from the $z\!\sim\!3$ LBG composite spectrum of \cite{shapley_rest-frame_2003}. This similarity suggests comparable relative contributions from neutral, low-ionization, and high-ionization gas phases in these two systems. In contrast, Galaxy~A ($z\simeq9.3$) spans a wider range in EW ratios. Lower-redshift samples show substantial object-to-object variation in these rest-frame ultraviolet absorption properties, with scatter of order a factor of two in ratios involving O\,I \citep[e.g.,][]{chisholm_shining_2016}. The range spanned by the present sample is therefore broadly consistent with the level of scatter seen at lower redshift, while also revealing clear variation among the three galaxies.

These trends show that galaxy-associated metal absorption at $z\gtrsim7$ exhibits significant system-to-system variation in ionization structure and/or kinematics. The observed diversity in EW ratios highlights the utility of EW-based diagnostics for probing early baryon cycling, particularly in regimes where line saturation, blending, and uncertain covering factors limit the robustness of abundance-based inferences (however, see \textit{Methods} and Extended Data Figs.~\ref{fig:NCIINCIV} and~\ref{fig:sio_co} for contextual comparison). The presence of chemically enriched gas traced by neutral, low-ionization, and high-ionization absorption by $z\sim9$ requires extremely rapid metal production, consistent with highly efficient early star formation. Recent theoretical work shows that such rapid enrichment can be achieved without invoking a dominant Population~III contribution, for example through Population~II star formation with a high upper-mass cutoff \citep[e.g.,][]{liu_impact_2025}, and is broadly consistent with expectations for very massive, low-metallicity stars in the early Universe \citep[e.g.,][]{jeong_simulating_2025,trinca_exploring_2024,fukushima_formation_2020}.

\subsection*{Implications for early baryon cycling during reionization.}
Taken together, these observations show that metal-enriched absorbing gas at $z\simeq7$--9 is associated with galaxies and exhibits non-thermal or unresolved kinematic structure. They indicate that key baryon-cycle ingredients, including metal enrichment and multiple ionic phases, were already present by the midpoint of reionization at $z\sim7$--8. The blueshifted absorption is consistent with outflowing gas, although turbulent interstellar motions or more complex internal gas kinematics may also contribute.

The host galaxies also show blue rest-frame ultraviolet continua ($\beta_{\rm UV,obs}=-2.19$, $-2.12$, and $-2.26$ for Galaxies~A, B, and C) with no evidence for extreme dust reddening. The SED fits favor sub-solar but non-negligible gas-phase metallicities, with $\log(Z_{\rm gas}/Z_\odot)=-0.52^{+0.09}_{-0.09}$, $-0.85^{+0.01}_{-0.01}$, and $-0.78^{+0.02}_{-0.02}$, respectively. These results indicate that detectable metal-enriched gas can coexist with relatively blue ultraviolet continua in at least some systems at these epochs, although a quantitative dust-to-metal constraint would require independent measurements of the total gas-phase metal and dust content. \rev{The blue ultraviolet continua and high luminosities of these galaxies also connect to physical scenarios proposed for ``blue monster'' galaxies, in which efficient early star formation, reduced effective dust attenuation, and feedback-driven redistribution of gas and dust shape the observed ultraviolet properties \citep{ziparo_blue_2023,ferrara_blue_2025}.} \rev{Similar physics may contribute to the blueshifted metal absorption observed here, supporting early feedback-driven enrichment and gas redistribution around luminous galaxies.}

The characteristic absorption offsets measured here, $\sim50$--250\,km\,s$^{-1}$, are broadly comparable to modest galaxy-associated gas flows reported in recent studies at high redshift, while remaining below the most extreme velocities seen in some AGN-dominated or strongly accelerated systems. Similar signatures have been reported from rest-frame ultraviolet absorption at $z\sim5$--6 \citep[e.g.,][]{sugahara_fast_2019}, from JWST-era ionized-gas outflow measurements at $z\sim2.5$--9 \citep[e.g.,][]{carniani_jades_2024,xu_shining_2025,cooper_high-velocity_2025,zhu_systematic_2025}, and from recent neutral- or Mg\,II-traced outflows in high-redshift quenching or stacked-galaxy samples \citep[e.g.,][]{wu_ejective_2025,valentino_gas_2025,lyu_first_2026,zhu_there_2026}; see also ALMA-based studies of outflowing gas in the early Universe \citep[e.g.,][]{gallerani_alma_2018,fujimoto_alpine-alma_2020}. Despite the different gas tracers and velocity definitions involved, these comparisons place the present measurements in the regime of modest galaxy-associated gas redistribution instead of unusually extreme feedback seen in quasars \citep[e.g.,][]{liu_frequent_2025}.

Continuum absorption-line measurements at $z>7$ are feasible only for the brightest galaxies with sufficiently high signal-to-noise rest-frame ultraviolet spectra. This criterion is met by three galaxies at $z\simeq7.2$--9.3 in SPURS, all of which show blueshifted neutral, low-ionization, and high-ionization metal absorption. These detections show that baryon-cycle signatures were already present in at least a subset of luminous galaxies by the midpoint of reionization.

At the same time, the observed diversity in EW ratios suggests that the physical state of this gas is not uniform across systems at these redshifts. While these results do not directly constrain the sources or timing of reionization, they argue against a picture in which galaxy environments remained chemically pristine until late times, and are consistent with emerging evidence that metal-enriched gas was already being redistributed prior to the completion of reionization \citep[e.g.,][]{davies_xqr-30_2023}.

% \clearpage

\section*{Methods}

\subsection*{Observations and data reduction.}
The data analyzed in this work were obtained as part of the JWST Cycle~4 program SPURS in the Abell 2744 field \citep[e.g.,][]{tang_spurs_2026, chen_spurs_2026}. 
Medium-resolution spectroscopy was carried out with JWST/NIRSpec \citep[e.g.,][]{jakobsen_near-infrared_2022} in MSA mode using the three medium-resolution gratings F100LP/G140M, F170LP/G235M, and F290LP/G395M, providing continuous wavelength coverage from $\sim1$ to $5\,\mu$m. Each galaxy was observed with all three gratings using a standard three-nod dithering pattern to improve background subtraction and mitigate detector artifacts. 
The effective exposure times were 29.2\,hr for G140M, 7.9\,hr for G235M, and 2.9\,hr for G395M.
In this work, we analyze the three galaxies at $z>7$ in SPURS whose rest-frame ultraviolet continuum near 1450\,\AA\ is detected at $S/N>10$, enabling absorption-line measurements against the stellar continuum.

The raw data were reduced and calibrated using the JWST Calibration Pipeline \citep[e.g.,][]{bushouse_jwst_2022}, version~1.14.0, with Calibration Reference Data System (CRDS) context \texttt{jwst\_1236.pmap}. The \texttt{calwebb\_detector1} stage was used to process the uncalibrated exposures and generate slope images, followed by the \texttt{calwebb\_spec2} stage to perform wavelength calibration, flat-fielding, flux calibration, and local background subtraction, producing rectified two-dimensional spectra for each nod position. Spectra from individual nods were then combined using the \texttt{calwebb\_spec3} stage. One-dimensional spectra were extracted using a boxcar aperture centered on the spatial trace of each galaxy. For all targets, spectra obtained with the three medium-resolution gratings were stitched together to form a continuous spectrum, with overlapping wavelength regions used to verify flux consistency. In addition to the standard pipeline steps, we applied custom procedures for hot pixel rejection, mitigation of low-level detector noise, and treatment of spatially extended sources, following the approach described in \cite{zhu_smiles_2026}. Flux uncertainties were propagated through all reduction steps and carried forward in subsequent measurements. We further scale the pipeline-provided flux uncertainties by a factor of 1.7 to account for residual pixel-to-pixel variations that are not fully captured by the formal error model \citep[e.g.,][]{brinchmann_high-z_2023,trump_physical_2023,maseda_jwstnirspec_2023}.

\subsection*{Photometric fitting and host-galaxy properties.}
To characterize the host galaxies, we fit their spectral energy distributions using publicly available HST and JWST imaging in the Abell~2744 field, fixing the redshift of each source to the NIRSpec spectroscopic measurement. The input photometry includes the HST filters F435W, F606W and F814W, together with JWST/NIRCam imaging spanning both broad and medium bands: F070W, F090W, F115W, F140M, F150W, F162M, F182M, F200W, F210M, F250M, F277W, F300M, F335M, F356W, F360M, F410M, F430M, F444W, F460M, and F480M.
We model these data with {\tt Prospector}, following the same fitting strategy as in \cite{zhu_smiles_2026}. The model includes flexible star-formation-history, stellar-population, dust-attenuation, and nebular-emission components appropriate for non-AGN star-forming galaxies at high redshift. 
\rev{We use a non-parametric SFH with seven age bins. The bin edges are logarithmically spaced from $\log_{10}(t/{\rm yr})=7.1295$ to $\log_{10}(0.9\,t_{\rm univ}/{\rm yr})$, with an additional final bin extending to $\log_{10}(t_{\rm univ}/{\rm yr})$, where $t_{\rm univ}$ is the age of the Universe at the galaxy redshift and $t$ is measured as lookback time from that redshift. In practice, for the redshift range of the three galaxies ($z=7.3$--$9.3$), the bin edges correspond to lookback times of approximately $0$, $24$--$26$, $44$--$49$, $80$--$94$, $144$--$179$, $261$--$341$, $471$--$651$, and $524$--$724$ Myr.}
For Galaxies~A, B, and C, the fits yield stellar masses of $\log(M_\star/M_\odot)=9.80^{+0.06}_{-0.05}$, $9.11^{+0.01}_{-0.01}$, and $9.48^{+0.01}_{-0.01}$, observed ultraviolet slopes of $\beta_{\rm UV,obs}=-2.19^{+0.01}_{-0.01}$, $-2.12^{+0.01}_{-0.01}$, and $-2.26^{+0.01}_{-0.01}$, dust attenuation parameters of ${\tt dust2}=0.45^{+0.08}_{-0.07}$, $1.13^{+0.02}_{-0.02}$, and $0.44^{+0.01}_{-0.01}$, stellar metallicities of $\log(Z_\star/Z_\odot)=-1.76^{+0.07}_{-0.09}$, $-1.68^{+0.02}_{-0.02}$, and \rev{$-1.978^{+0.002}_{-0.001}$}, and gas metallicities of $\log(Z_{\rm gas}/Z_\odot)=-0.52^{+0.09}_{-0.09}$, $-0.85^{+0.01}_{-0.01}$, and $-0.78^{+0.02}_{-0.02}$, respectively. \rev{The uncertainties quoted here correspond to the 16th--84th percentile ranges of the posterior distributions; the typical systematic uncertainty of this SED-fitting framework is $\pm0.15$ dex \citep{zhu_smiles_2026}.}
The recent star-formation histories inferred from the fits are also substantial, with youngest-bin star-formation rates of 19, 39, and $29\,M_\odot\,{\rm yr^{-1}}$ for Galaxies~A, B, and C, respectively. 

\subsection*{Systemic redshifts and velocity reference frame.}
Accurate systemic redshifts are essential for interpreting absorption-line kinematics. \rev{Extended Data Fig.~\ref{fig:ed_nebular_lines} shows observed-frame cutouts of H$\beta$ $\lambda4861$ and the [O\,III] $\lambda\lambda4959,5008$ doublet for all three galaxies.}
For each galaxy, we adopt the centroid of [O\,III] $\lambda5008$ as the fiducial systemic reference, as this is the strongest and most robustly measured nebular line in the current data. We independently fit H$\beta$ where detected and use it as a consistency check on the systemic frame. 
% For Galaxy~C, H$\beta$ agrees with [O\,III] $\lambda5008$ within $\sim5$\,km\,s$^{-1}$, whereas for Galaxy~B it is offset redward by $\sim90$\,km\,s$^{-1}$; adopting H$\beta$ in that system increases the inferred absorption blueshift. For Galaxy~A, H$\beta$ is only weakly detected and does not provide a comparably reliable independent centroid. Emission-line centroids were measured by fitting Gaussian profiles to continuum-subtracted spectra, with uncertainties estimated from the covariance matrix of the fit. 
For Galaxy~C, H$\beta$ agrees with [O\,III] $\lambda5008$ within $\sim5$\,km\,s$^{-1}$, whereas for Galaxy~B it is offset redward by $\sim90$\,km\,s$^{-1}$; adopting H$\beta$ in that system increases the inferred absorption blueshift. For Galaxy~A, H$\beta$ is only weakly detected and does not provide a comparably reliable independent centroid. \rev{Lower-redshift interloper solutions are disfavored by the combination of the Ly$\alpha$ break, the H$\beta$+[O\,III] line pattern at a common redshift, and the absence of other sources within the MSA slitlets.} Emission-line centroids were measured by fitting Gaussian profiles to continuum-subtracted spectra, with uncertainties estimated from the covariance matrix of the fit.
\rev{The nebular profiles show no separate, well-constrained broad or shifted emission component.} \rev{We therefore use the nebular lines to define the systemic frame, while the absorption-line kinematics provide the evidence for outflowing or otherwise disturbed gas.}
For Galaxy~A, [O\,III] $\lambda5008$ lies near the red end of the G395M wavelength range, but the fitted line window remains within the available spectral coverage, the wavelength solution extends beyond the fitted window, and the line centroid is not clipped by the data edge. All absorption-line velocities are computed relative to these adopted nebular systemic redshifts, and Table~\ref{tab:ed_abs}  reports the corresponding velocity uncertainties. The absorption-line velocities are reported relative to this systemic frame as
\begin{equation}
    v = c\,\frac{z_{\rm abs}-z_{\rm sys}}{1+z_{\rm sys}},
\end{equation}
where $z_{\rm abs}$ is the absorber redshift and $c$ is the speed of light. We also considered whether the stellar continuum could provide an independent systemic velocity reference, but the spectra do not show stellar absorption features strong enough to yield a velocity zero point with precision comparable to the nebular-line centroids.

For completeness, we also measure rest-frame optical nebular emission-line flux ratios from the same NIRSpec spectra used to determine systemic redshifts. The measured [O\,III] $\lambda5008$/H$\beta$ ratios are $8.0 \pm 1.7$, $8.1 \pm 0.3$, and $6.7 \pm 0.2$ for Galaxies~A, B, and C, respectively. Such high ratios are commonly observed in low-metallicity star-forming galaxies at high redshift and can also arise under harder ionizing conditions. We do not attempt to use these line ratios to discriminate between ionizing sources or excitation mechanisms. 
Accordingly, the current spectra do not show a definitive AGN diagnostic, such as broad emission lines with FWHM$>1000\rm \, km\,s^{-1}$.
Importantly, the identification, kinematics, and abundance inferences of the absorption features presented in this work do not depend on the physical origin of the nebular emission. 
% The nebular lines are used solely to establish accurate systemic redshifts, which define the velocity reference frame for the absorption analysis.

\subsection*{Continuum normalization, absorption-line identification, and profile fitting.}
We measure absorption-line properties in a two-step procedure that (i) fits a local linear continuum around each transition and produces a continuum-normalized spectrum, and (ii) fits Voigt absorption models to the normalized profiles. For each transition, we extract a wavelength window of typically $\pm300$--500\,\AA\ around the expected observed wavelength based on $z_{\rm sys}$. We fit the local continuum with a linear model, $f_\lambda = a + b(\lambda-\lambda_0)$, using inverse-variance weighting. To avoid bias from line features, we mask a central region of $\pm12$--25\,\AA\ around the expected line center (depending on transition and spectral resolution) and iteratively sigma-clip ($4\sigma$) outliers in the remaining continuum region repeated for up to three iterations. In addition, when other obvious absorption features fall within the continuum window (e.g., unrelated lower-redshift absorbers not discussed in this work), we manually mask those regions prior to the continuum fit; for the Galaxy~C C\,IV doublet, this includes masking the adjacent blue-side absorption feature before fitting the local continuum. Masked intervals due to noisy data are visible as locally increased uncertainties in the relevant wavelength ranges. The observed spectrum is divided by the best-fitting linear continuum to obtain a continuum-normalized flux and uncertainty spectrum.

Absorption features are identified at the expected wavelengths of known metal transitions, requiring $\gtrsim3\sigma$ significance, where available, consistent detections across multiple transitions tracing the same ion or ionization phase. We fit the continuum-normalized absorption profiles using Voigt models. For isolated single transitions (O\,I $\lambda1302$, Si\,II $\lambda1260$, C\,II $\lambda1334$), we fit a single component characterized by the absorber redshift $z_{\rm abs}$, a Gaussian width $\sigma_g$ (in wavelength units), a Lorentzian damping parameter $\gamma_l$, and a line-center optical-depth normalization $\tau_0$ of an area-normalized Voigt profile. Fits are performed via $\chi^2$ minimization, with parameter bounds chosen to avoid unphysical solutions, including enforcing a minimum resolved width comparable to the instrumental core. For resonance doublets (Si\,IV $\lambda\lambda1393,1402$ and C\,IV $\lambda\lambda1548,1550$), the two components are fit simultaneously, enforcing a common absorber redshift and kinematic width and fixing the optical-depth ratio to the ratio of oscillator strengths. Equivalent widths for the two components are measured separately within integration windows defined as $\pm3\sigma_g$ around each fitted line center, with the additional requirement that the integration domains lie on opposite sides of the midpoint between the two components to prevent overlap when the doublet is partially blended.

At the spectral resolution of these data ($R\sim1000$), the fitted profiles and measured equivalent widths should be interpreted as effective, unresolved representations of potentially multiple narrow components, following e.g., \cite{shapley_rest-frame_2003}. In particular, for C\,II $\lambda1334$, the measured equivalent width may include contributions from unresolved fine-structure absorption (C\,II$^\ast$ $\lambda1335$) and weak wing structure, which cannot be reliably separated at the present resolution and signal-to-noise. We also identify tentative absorption consistent with Si\,II$^\ast$ $\lambda1264$ in Galaxy~A (e.g., \cite{ding_deep_2020}); however, given the low signal-to-noise and potential blending with the Si\,II $\lambda1260$ profile, we do not attempt an explicit fit or equivalent-width measurement for this feature.

We report two equivalent-width estimates for each transition. The primary measurement is the data equivalent width, computed by direct integration of $(1-F_{\rm norm})$ over an objectively defined line window. For single transitions, the integration window is taken as $\pm3\sigma_g$ around the fitted line center. For doublets, equivalent widths are computed for each component within $\pm3\sigma_g$ of its fitted center (split at the doublet midpoint) and summed to obtain a total doublet equivalent width. The corresponding statistical uncertainty is computed by propagating the normalized flux uncertainties across the same integration window. In addition, we compute a model equivalent width by integrating the best-fitting Voigt model over the same window; its uncertainty is estimated from the parameter covariance via finite-difference propagation. Rest-frame equivalent widths are obtained by dividing observed-frame values by $(1+z_{\rm abs})$. We also record the minimum continuum-normalized flux within the integration window as a model-independent measure of line depth and saturation. The fitted optical-depth normalizations $\tau_0$ are provided for reference only. Given the moderate spectral resolution, the presence of saturation, and the possibility of partial covering, $\tau_0$ and Voigt-derived column densities are not used as primary constraints. Throughout the main text, we therefore emphasize equivalent widths, minimum normalized fluxes, and overlapping kinematics as the most robust observables. Our measurements are summarized in Table \ref{tab:ed_abs}.

Extended Data Fig.~\ref{fig:ed_abs_validation} presents a validation of these detections for all three galaxies in this work. For each galaxy and transition, the two-dimensional NIRSpec spectra and the corresponding one-dimensional extracted spectra with the best-fitting absorption profiles overlaid are shown, together with the integration windows used for equivalent-width measurements after local continuum normalization. The spatial coincidence of the absorption features with the galaxy continuum trace in the two-dimensional spectra shows that the detected lines are associated with the target galaxies instead of detector artifacts or background residuals. \rev{The fitted line centroids also show blueshifted absorption and ionization-dependent velocity structure, summarized in Fig.~\ref{fig:ion_velocity}.}

\subsection*{Line centroid uncertainty estimation.}
To quantify realistic uncertainties on line centroid measurements given the moderate spectral resolution of the NIRSpec medium--resolution gratings ($R \simeq 1000$), we performed mock line-injection tests directly on the observed, calibrated one-dimensional spectrum. Synthetic Gaussian absorption features were injected at representative wavelengths (e.g., $\lambda \simeq 15{,}500~\text{\AA}$) with widths fixed by the typical grating resolution, $\sigma = \lambda/(2.355\,R)$, and with depths comparable to those of the detected absorption lines. The injected features were added to the real spectrum, preserving the native wavelength grid, noise properties, and any correlated residual structure. For each realization, we refit the line centroid using the same fitting procedure applied to the science data. Repeating this procedure for 3000 Monte Carlo realizations, we find a negligible centroid bias ($\simeq -15~\mathrm{km~s^{-1}}$) and a $1\sigma$ scatter of $\simeq 47~\mathrm{km~s^{-1}}$.

As discussed earlier, the formal pipeline uncertainties underestimate the true pixel-to-pixel variance in the NIRSpec spectra. When the flux uncertainties are conservatively scaled by a factor of 1.7 to account for this effect, the corresponding centroid scatter increases to $\sim133~\mathrm{km~s^{-1}}$. The velocity uncertainties reported in Table~\ref{tab:ed_abs} are the statistical fit uncertainties from the profile fits; the injection-based scatter is used as a conservative guide to the robustness of individual centroid offsets. The repeated occurrence of blueshifted absorption across multiple transitions and galaxies, together with the overlapping velocity structure within each system, indicates that the measured velocity shifts are not driven by a single marginal centroid measurement. The injection tests provide an empirical estimate of the typical centroid uncertainty.

\subsection*{Constraints on gas temperature from line widths.}
As a consistency check, we use the measured absorption-line widths to place thermal-only upper limits on the gas temperature. For each transition, we adopt the fitted Gaussian dispersion in wavelength units, $\sigma_g$, from the profile fitting described above, and convert it to a velocity dispersion using a reference wavelength $\lambda_{\rm ref}$ (the fitted line center for single transitions and the midpoint of the fitted doublet components for Si\,IV and C\,IV).

Because the observed widths include instrumental broadening, we correct them using a Gaussian approximation to the instrumental profile with resolving power $R\simeq1000$. We take ${\rm FWHM}_{\rm inst}=\lambda_{\rm ref}/R$ and $\sigma_{\rm inst}={\rm FWHM}_{\rm inst}/2.355$, and estimate the intrinsic width via quadrature subtraction, $\sigma_{\rm int}^2=\sigma_g^2-\sigma_{\rm inst}^2$. This correction is intended only for order-of-magnitude thermal checks. The exact NIRSpec line-spread function varies with wavelength, so $\sigma_{\rm int}$ values near the instrumental limit should not be over-interpreted. At the same time, because the effective resolution is generally better at the longer observed wavelengths corresponding to higher-redshift lines, the resolved/unresolved classification is unlikely to be set primarily by this wavelength dependence. Transitions that are unresolved under this correction (i.e., $\sigma_g \lesssim \sigma_{\rm inst}$ within a small tolerance) or that reach the imposed minimum width in the fitting procedure are excluded from the temperature analysis. Accordingly, differences in fitted $\sigma_{\rm int}$ among transitions do not by themselves imply distinct bulk kinematics, as they can also reflect unresolved substructure, saturation, blending, and measurement uncertainty near the resolution limit.

We convert $\sigma_{\rm int}$ to a Doppler parameter $b=c\,\sigma_{\rm int}/\lambda_{\rm ref}$ and compute
\begin{equation}
    T_{\max} = \frac{m\,b^2}{2k_{\rm B}},
\end{equation}
where $m$ is the atomic mass of the ion and $k_{\rm B}$ is the Boltzmann constant. These values are upper limits on the thermal temperature, since any contribution from turbulence, bulk flows, or unresolved velocity substructure would reduce the thermal component of the line width. The resulting limits are therefore interpreted only as indicative checks on physical plausibility instead of direct temperature measurements.

The inferred $T_{\max}$ values for neutral (O\,I), low-ionization (Si\,II, C\,II), and high-ionization (Si\,IV, C\,IV) species in each galaxy are shown in Extended Data Fig.~\ref{fig:tmax}. Because these limits exceed the ionization survival temperatures of the detected species, the observed line widths are likely dominated by non-thermal motions. This is consistent with lower-redshift studies in which multiphase galaxy-associated absorption reflects substantial unresolved or turbulent kinematic structure \citep[e.g.,][]{churchill_low-_2000,li_kinematically_2026,sun_census_2026}.

\subsection*{Ionic column-density lower limits and relative carbon absorption.}

\rev{We first compute conservative ionic column-density lower limits from the measured rest-frame equivalent widths. We use the optically thin linear curve-of-growth relation \citep{draine_physics_2011}}
\begin{equation}
    \rev{N_{\rm ion}=1.13\times10^{20}\frac{W_0}{\sum_i f_i\lambda_i^2}\ {\rm cm}^{-2},}
\end{equation}
\rev{where $W_0$ and $\lambda_i$ are in \AA, and the sum is taken over both components for doublets. These values, listed in Extended Data Table~\ref{tab:ed_column_limits}, are lower limits because unresolved saturation and partial covering can increase the true ionic columns. We do not infer a total metal mass or $M_{\rm metal}/M_\star$, since that conversion requires the absorber area, covering fraction, geometry, and ionization correction.}

\rev{The optically thin limits in Extended Data Table~\ref{tab:ed_column_limits} are intended only as conservative ionic column-density lower limits and are not used to derive the carbon ratios shown in Extended Data Fig.~\ref{fig:NCIINCIV}.} As a qualitative, order-of-magnitude probe of the ionization structure of the galaxy-associated absorbing gas, we examine the ratio of low- to high-ionization carbon, $N(\mathrm{C\,II})/N(\mathrm{C\,IV})$, for systems with coverage of both transitions. Ratios between C\,II and C\,IV have been widely used as empirical indicators of multiphase gas in absorption systems; however, their interpretation depends sensitively on line saturation, covering fraction, and spectral resolution. At the moderate resolution of the NIRSpec medium-resolution gratings ($R\sim1000$), intrinsically saturated absorption with partial covering can appear unsaturated, placing the inferred column densities on the flat part of the curve of growth and allowing order-of-magnitude variations in $N$ for modest changes in line depth.

\rev{For the carbon-ratio comparison, we use optical-depth integrals of the best-fitting Voigt profiles.} For C\,II $\lambda1334$, we adopt the fitted $(\tau_0,\sigma_g)$ from the single-line fit. For the C\,IV doublet, we use the $(\tau_0,\sigma_g)$ from the simultaneous doublet fit and adopt an effective oscillator strength equal to the sum of the two components. Given the likelihood of unresolved saturation and non-uniform covering, the resulting ratios should not be interpreted as precise column-density measurements.

As shown in Extended Data Fig.~\ref{fig:NCIINCIV}, the inferred $N(\mathrm{C\,II})/N(\mathrm{C\,IV})$ ratios fall within the broad range spanned by quasar-selected absorbers at $z\sim2$--6 \citep[e.g.,][]{boksenberg_properties_2015,rowlands_e-xqr-30_2026}, including systems at $5\lesssim z\lesssim6.3$ from the E-XQR-30 sample \citep[e.g.,][]{rowlands_e-xqr-30_2026}; also see \cite{becker_evolution_2019,becker_iron_2012,becker_reionisation_2015,christensen_metal_2023,cooper_heavy_2019,davies_examining_2023,davies_xqr-30_2022,davies_xqr-30_2023,dodorico_evolution_2022,dodorico_xqr-30_2023,doughty_evolution_2019,fan_quasars_2023,noterdaeme_proximate_2019,sodini_evidence_2024,wu_c_2021,zou_spectroscopic_2024}. We use this comparison strictly as a consistency check, noting that the present data do not permit robust constraints on ionization state from carbon ratios alone.

To provide qualitative guidance on the ionization conditions compatible with the observed ratios, we computed a set of simple photoionization models using \textsc{Cloudy} (c25; \citealp{gunasekera_2025_2025}). The models assume plane-parallel geometry, constant density, and a single-phase slab illuminated by an external radiation field. We explored a broad range of ionization parameters, gas densities, and total hydrogen column densities, focusing on reproducing the observed $N(\mathrm{C\,II})/N(\mathrm{C\,IV})$ ratios. Matching the observed values requires ionization parameters of $\log U \simeq -1.9$ to $-2.0$ across a wide range of densities. These models are intended to provide qualitative guidance only; our main conclusions do not depend on the details of the photoionization modeling and instead rest on the overlapping kinematics and coexistence of low- and high-ionization absorption.

\subsection*{Relative abundance ratios}
We also compute approximate coordinates in the $[\mathrm{Si/O}]$--$[\mathrm{C/O}]$ plane (Fig.~\ref{fig:sio_co}) using the neutral transition O\,I $\lambda1302$ and the low-ionization transitions C\,II $\lambda1334$ and Si\,II $\lambda1260$. For each line, we adopt the model rest-frame equivalent width $W_{0,\rm model}$ obtained from the best-fitting Voigt profile and convert to an optically thin column density using the linear curve-of-growth relation. Solar abundance ratios are taken from \cite{asplund_chemical_2009}.

At the spectral resolution of these data, the low-ionization transitions are strong and likely affected by unresolved saturation and non-uniform covering fractions. In this regime, modest changes in optical depth or ionization corrections can lead to substantial shifts in inferred column densities. The resulting $[\mathrm{Si/O}]$ and $[\mathrm{C/O}]$ values should therefore be interpreted as empirical line-strength ratios mapped onto abundance space for rough comparison.

For reference, we compare these systems to quasar-selected absorbers from the XQR-30 compilation \citep{sodini_evidence_2024} and to representative chemical-evolution model regions from \cite{vanni_chemical_2024}, as implemented by \cite{sodini_evidence_2024}. We also add quasar-selected absorbers from JWST/NIRSpec observations in \cite{christensen_metal_2023} for reference. We do not attempt to infer nucleosynthetic channels or stellar populations from these comparisons. Instead, we use the diagram as a qualitative consistency check: all three galaxies occupy regions overlapping previously studied high-redshift absorbers and do not exhibit extreme offsets relative to lower-redshift systems. Within the systematic uncertainties described above, the line ratios are therefore consistent with enrichment patterns commonly observed in metal-bearing gas at later cosmic times.

At the same time, the presence of chemically enriched gas associated with galaxies by $z\sim9$ places strong constraints on the timescale and efficiency of early metal production \citep[e.g.,][]{wise_birth_2012,wise_birth_2014,hafen_origins_2019,kim_high-redshift_2019}. These observations suggest that substantial carbon and oxygen enrichment can arise extremely rapidly, as illustrated by Galaxy~A at $z=9.3$ in this work and by the detection of strong rest-frame optical C and O emission lines in galaxies at $z>14$ \citep[e.g.,][]{carniani_eventful_2025,naidu_cosmic_2025,helton_ionizing_2025}. Recent models show that such rapid enrichment does not require a dominant Population~III contribution: Population~II star formation with an upper stellar mass cutoff of $\sim200$--$300\,M_\odot$ can efficiently reproduce high metal yields on short timescales \citep[e.g.,][]{liu_impact_2025}. Similar very massive, low-metallicity stars have been invoked to explain the high ultraviolet luminosities of galaxies at $z\gtrsim10$ \citep[e.g.,][]{jeong_simulating_2025,trinca_exploring_2024}, consistent with theoretical expectations that the maximum stellar mass increases at low metallicity due to reduced opacity and enhanced radiative transparency \citep[e.g.,][]{omukai_low-metallicity_2008,fukushima_formation_2020}. In this sense, while the relative-abundance measurements presented here do not obviously require exotic enrichment channels, they are consistent with an early onset of baryon cycling driven by highly efficient star formation in the first galaxies.

\section*{Data Availability}
This work is based on observations made with the NASA/ESA/CSA James Webb Space Telescope. The data were obtained from the Mikulski Archive for Space Telescopes at the Space Telescope Science Institute, which is operated by the Association of Universities for Research in Astronomy, Inc., under NASA contract NAS 5-03127 for JWST. These observations are associated with program \# 9214. The data described here may be obtained from the MAST archive at 
\url{https://dx.doi.org/10.17909/qz3x-3868}.

\section*{Code Availability}
Analysis was performed using \texttt{Astropy} \citep{astropy_collaboration_astropy_2022}, 
\texttt{CLOUDY} \citep{gunasekera_2025_2025}, the \texttt{JWST Calibration Pipeline} \citep{bushouse_jwst_2022},
\texttt{NumPy} \citep{van_der_walt_numpy_2011}, and \texttt{Matplotlib} \citep{hunter_matplotlib_2007}. 

\section*{Correspondence}
Correspondence and requests for materials should be addressed to Yongda Zhu and Zhiyuan Ji.

\section*{Acknowledgments}
We thank the anonymous reviewers for their helpful comments.
The authors acknowledge the SPURS Team (PID-9214, PIs: Charlotte Mason, Dan Stark; CoIs: Zuyi Chen et al.) for developing their observing program with a zero-exclusive-access period. YZ and ZJ acknowledge support from the NIRCam Science Team contract to the University of Arizona, NAS5-02105. YZ is also supported by JWST Program \#6434. Support for program \#6434 was provided by NASA through a grant from the Space Telescope Science Institute, which is operated by the Association of Universities for Research in Astronomy, Inc., under NASA contract NAS 5-03127.

We respectfully acknowledge the University of Arizona is on the land and territories of Indigenous peoples. Today, Arizona is home to 22 federally recognized tribes, with Tucson being home to the O’odham and the Yaqui. The university strives to build sustainable relationships with sovereign Native Nations and Indigenous communities through education offerings, partnerships, and community service.

ChatGPT \citep{openai_chatgpt_2024} was used for language editing of this manuscript. The authors are responsible for all scientific content.

\section*{Author Contributions}
Y.Z. and Z.J. initiated and led the project.  
Y.Z. led the data analysis and wrote the initial draft of the manuscript.  
Z.J. contributed to target selection, interpretation of the results, and scientific discussion.  
G.D.B., E.E., X.F., G.H.R., and M.J.R. contributed to the scientific framing and presentation of the results.  
J.D., X.J., and Z.M. contributed to the identification and verification of absorption features.  
W.L., J.L., S.N., Y.W., M.Y., and J.Z. provided comments and discussions that improved the manuscript.

\section*{Competing interests}
The authors declare no competing interests.

% References
% \bibliographystyle{astjnlabbrev-nature}
% \bibliography{references}{}

\clearpage

\setcounter{figure}{0}
\renewcommand{\figurename}{Extended Data Figure}
\setcounter{table}{0}
\renewcommand{\tablename}{Extended Data Table}

\begin{figure*}[!ht]
    \centering
    \includegraphics[width=1\linewidth]{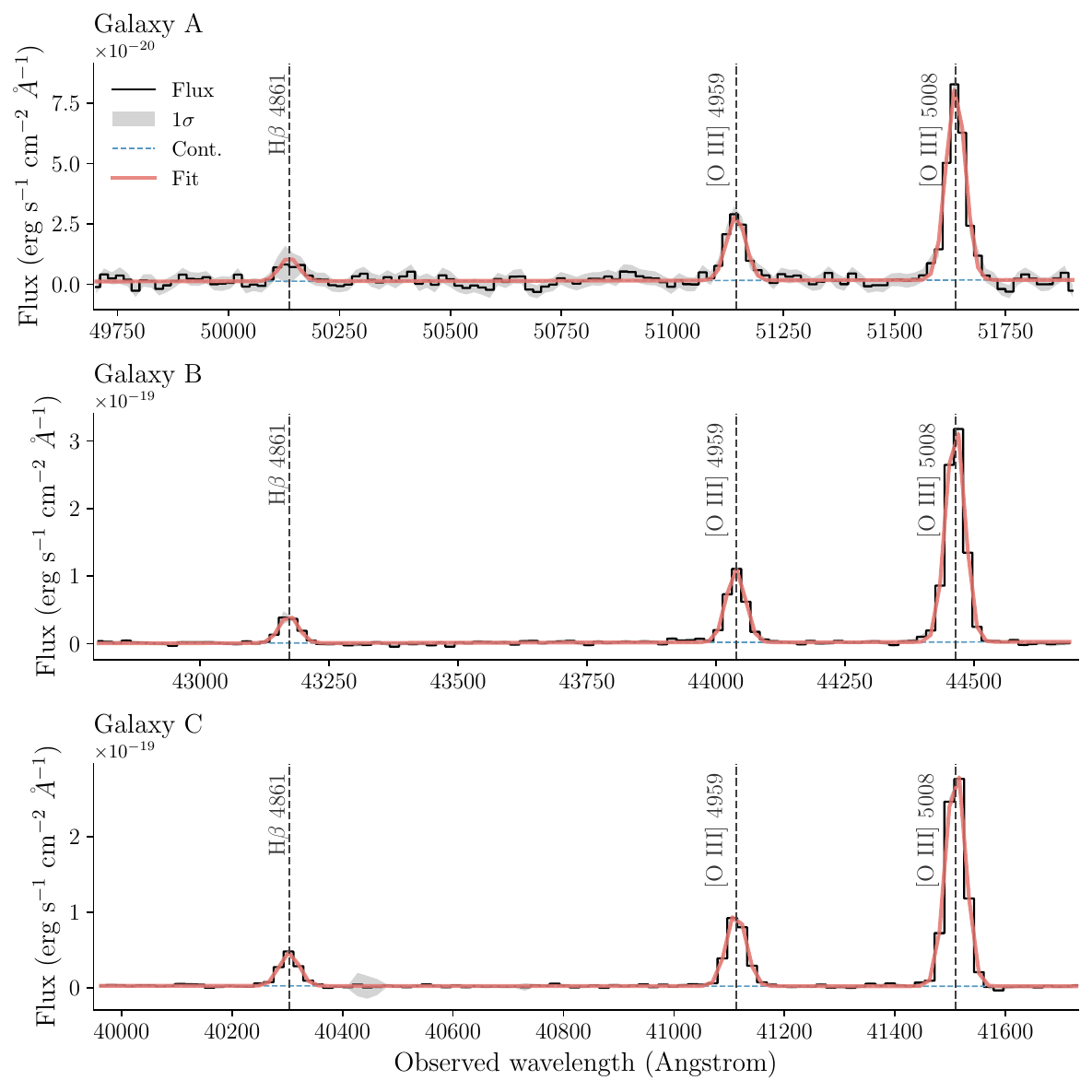}
    \caption{\rev{Observed-frame nebular-line cutouts for the three galaxies. Each panel shows H$\beta$ $\lambda4861$ and the [O\,III] $\lambda\lambda4959,5008$ doublet. Black curves show the observed NIRSpec spectra, grey shading shows the $1\sigma$ uncertainty, blue dashed curves show the local continuum, and red curves show Gaussian emission-line fits. Vertical dashed lines mark the expected observed wavelengths of H$\beta$, [O\,III] $\lambda4959$, and [O\,III] $\lambda5008$ based on the adopted systemic redshift.}
}
\label{fig:ed_nebular_lines}
\end{figure*}

\begin{figure*}[!ht]
\centering
\includegraphics[width=0.95\textwidth]{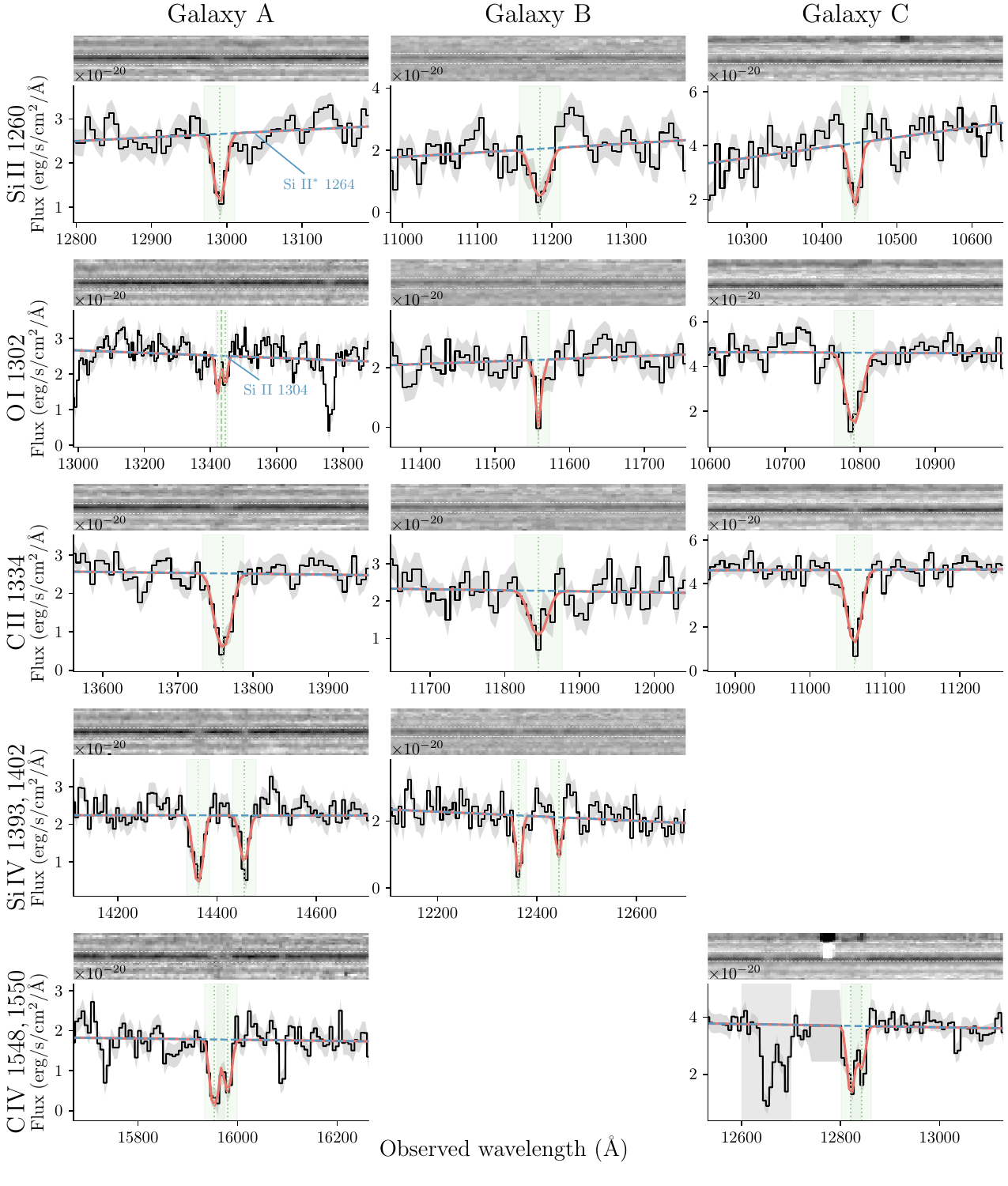}
\caption{
\rev{Validation of absorption-line detections. Columns show Galaxies~A, B, and C from left to right. Rows show Si\,II $\lambda1260$, O\,I $\lambda1302$(+Si\,II $\lambda1304$), C\,II $\lambda1334$, Si\,IV $\lambda\lambda1393,1402$, and C\,IV $\lambda\lambda1548,1550$ from top to bottom. In each panel, the upper sub-panel shows the two-dimensional NIRSpec spectrum, and the lower sub-panel shows the extracted one-dimensional spectrum with the best-fitting absorption profile. Grey shading marks the $1\sigma$ flux uncertainty, and vertical shaded regions mark the equivalent-width integration windows. Galaxy~A also shows tentative absorption consistent with Si\,II$^\ast$ $\lambda1264$.}
}
\label{fig:ed_abs_validation}
\end{figure*}

\begin{figure*}[!ht]
\centering
\includegraphics[width=0.6\linewidth]{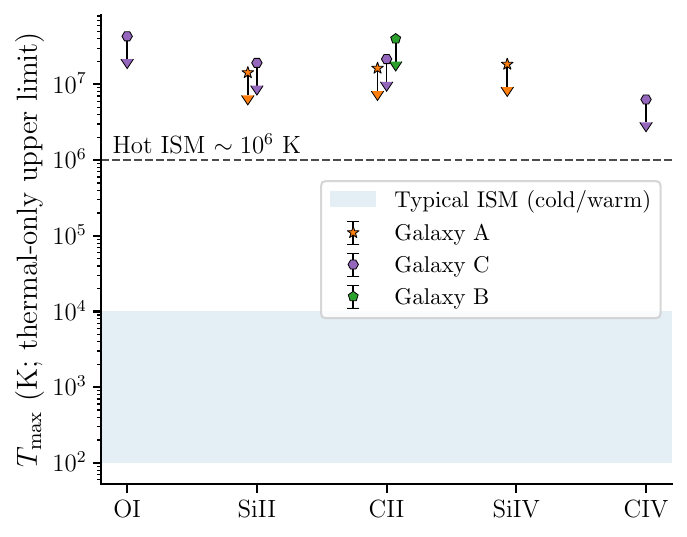}
\caption{
Thermal-only upper limits on the gas temperature inferred from the measured velocity dispersions of metal absorption lines. For each galaxy, we show the maximum temperature $T_{\max}$ implied by the fitted Gaussian core width under the assumption that the observed line width arises purely from thermal broadening, for neutral (O\,I), low-ionization (Si\,II, C\,II), and high-ionization (Si\,IV, C\,IV) species. Because any contribution from turbulence, bulk motions, or unresolved velocity substructure would reduce the thermal component of the line width, the inferred values represent strict upper limits on the gas temperature. Shaded regions indicate typical temperature ranges of cold and warm interstellar gas ($T\sim10^{2}$--$10^{4}$\,K), shown for reference.
}
\label{fig:tmax}
\end{figure*}

\begingroup
\begin{table*}[!ht]
    \centering
    % \fontsize{8pt}{9pt}\selectfont
    \small 
\begin{tabular}{lccc}
\toprule
 & Galaxy A & Galaxy B & Galaxy C \\
\midrule
O\,I 1302 & $15.16\pm0.07$ & $15.19\pm0.11$ & $15.47\pm0.04$ \\
Si\,II 1260 & $13.82\pm0.07$ & $14.14\pm0.09$ & $13.82\pm0.08$ \\
C\,II 1334 & $15.00\pm0.04$ & $14.93\pm0.10$ & $15.01\pm0.04$ \\
Si\,IV 1393,1402 & $14.36\pm0.04$ & $14.22\pm0.07$ & \dots \\
C\,IV 1548,1550 & $14.75\pm0.04$ & \dots & $14.59\pm0.04$ \\
\bottomrule
\end{tabular}
\caption{
\rev{Conservative ionic column-density lower limits, reported as $\log_{10}(N_{\rm ion}^{\rm thin}/{\rm cm}^{-2})$.}
}
\label{tab:ed_column_limits}
\end{table*}
\endgroup

\begin{figure*}[!ht]
\centering
\includegraphics[width=0.7\linewidth]{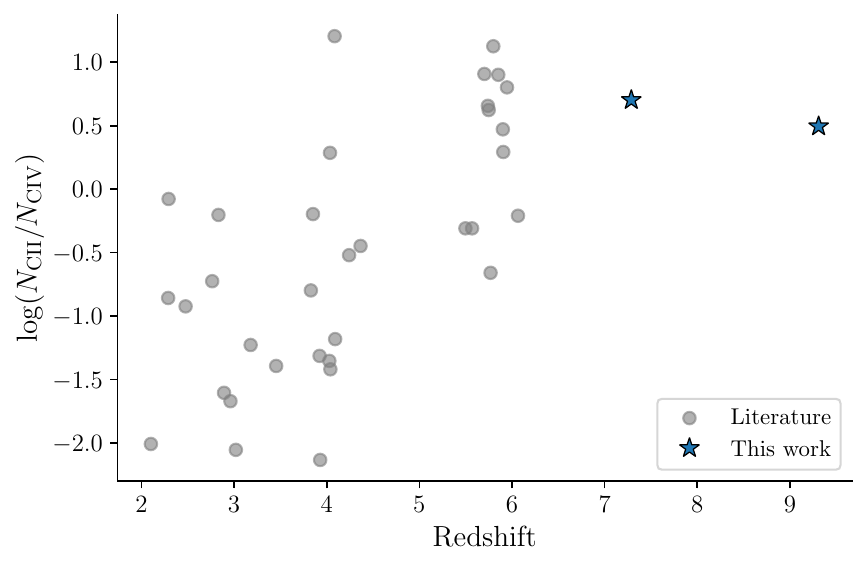}
\caption{
\rev{Approximate ratio of low- to high-ionization carbon column densities, $\log(N_{\mathrm{C\,II}}/N_{\mathrm{C\,IV}})$, as a function of redshift.} Grey circles show literature measurements of quasar-selected absorbers at $z\sim2-6$, including $z<5$ systems from \cite{boksenberg_properties_2015} and $z>5$ systems from \cite{rowlands_e-xqr-30_2026}. Blue stars mark the galaxy-associated absorbers analyzed in this work at $z\simeq7.3$ and $z\simeq9.3$, for systems with coverage of both C\,II and C\,IV. The values measured here fall within the broad range spanned by absorbers at $5\lesssim z\lesssim 6.3$ in \cite{rowlands_e-xqr-30_2026}. Given the limited sample size and heterogeneous selection of the comparison data, no inference on redshift evolution is made.
}
\label{fig:NCIINCIV}
\end{figure*}

\begin{figure*}[!ht]
\centering
\includegraphics[width=0.7\linewidth]{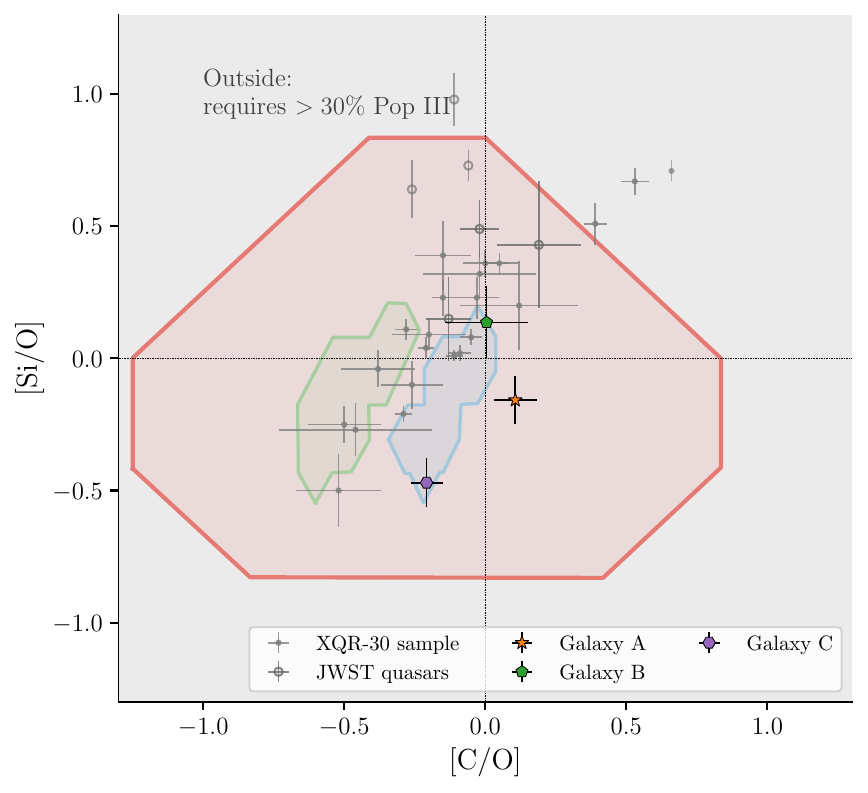}
\caption{
Relative abundance ratios $[\mathrm{Si/O}]$ and $[\mathrm{C/O}]$ for galaxy-associated absorbers at $z=7.2$--$9.3$, compared with quasar-selected absorbers and simplified chemical-enrichment model regions. Grey symbols show the XQR-30 sample of $z\gtrsim5$ damped and proximate damped Ly$\alpha$ systems along quasar sightlines \cite{sodini_evidence_2024}, with error bars indicating $1\sigma$ uncertainties, while open grey symbols denote JWST/NIRSpec quasar-selected absorbers at $z>6.5$ from \cite{christensen_metal_2023}. Coloured symbols mark the three galaxies analyzed in this work: Galaxy~A at $z=9.31$ (star), and Galaxies~B and~C at $z=7.88$ and $z=7.29$ (pentagon and hexagon, respectively). Shaded model regions illustrate relative-abundance parameter space predicted by chemical evolution models \citep[e.g.,][]{vanni_chemical_2024}, as implemented in \cite{sodini_evidence_2024}: the inner regions (green and blue) correspond to negligible Population~III contributions ($\lesssim0.01\%$; \cite{limongi_presupernova_2018} and \cite{woosley_evolution_1995}), while the red enclosing boundary marks the locus consistent with non-dominant Population~III enrichment ($<30\%$), based on nucleosynthetic yields. The model region is shown for $[\mathrm{O/H}]>-4$.
}
\label{fig:sio_co}
\end{figure*}

\end{document}